\documentclass[preprint,preprintnumbers,nofootinbib,aps,10pt]{revtex4-1}
\usepackage{amsmath,amssymb,bm,epsfig}
\usepackage{color}
\usepackage{natbib}
\usepackage{hyperref}
\usepackage{ulem}
\usepackage{graphicx}
\usepackage{xcolor,colortbl}
\usepackage{pifont}

\def \beq{\begin{equation}}
\def \eeq{\end{equation}}
\def \beqa{\begin{eqnarray}}
\def \eeqa{\end{eqnarray}}
\def \l{\left(}
\def \r{\right)}

\usepackage{xspace}

\newcommand{\cmark}{\ding{51}}%
\newcommand{\xmark}{\ding{55}}%

\makeatletter
\newcommand{\thickhline}{%
    \noalign {\ifnum 0=`}\fi \hrule height 1pt
    \futurelet \reserved@a \@xhline
}
\newcolumntype{"}{@{\hskip\tabcolsep\vrule width 1pt\hskip\tabcolsep}}
\makeatother

\definecolor{Gray}{gray}{0.87}
\definecolor{LightCyan}{rgb}{0.88,1,1}
\definecolor{lmv}{rgb}{0.86, 0.82, 1.0}

\begin{document}

\preprint{MSUHEP-150202}

\title{Sign Structure of Susceptibilities of Conserved Charges in the 
 $\l 2+1\r$ Polyakov Quark Meson Model}
 
\author{Sandeep Chatterjee}
\email{sandeepc@vecc.gov.in}
\affiliation{Theoretical Physics Division, Variable Energy 
Cyclotron Centre, Kolkata 700064, India}

 \author{Kirtimaan A. Mohan}
\email{kamohan@pa.msu.edu}
\affiliation{Department of Physics and Astronomy, Michigan State University, East Lansing, Michigan 48824, USA}


\begin{abstract}
The sign structure of correlations of conserved charges are investigated in a QCD like 
model: the $\l2+1\r$ flavor Polyakov Quark Meson model. We compute all 
susceptibilities of the conserved charges on the $\l\mu_B-T\r$ plane up to 
fourth order and a few at higher order as well. By varying the mass of the sigma meson, 
we are able to study and compare scenarios with as well as without a critical point. 
In the hadron-quark transition regime we identify certain 
correlations that turn negative unlike expectation from ideal hadron resonance gas 
calculations. The striking feature being that these remain negative deep into the hadronic side and thus could be measured in experiments. Measurement of such quantities in heavy ion collision 
experiments can elucidate the location of the QCD transition curve and possibly the critical point. 
\end{abstract}
\maketitle

The phase diagram of Quantum Chromodynamics (QCD), the theory of strongly interacting matter, 
has been a subject of intense study both theoretically and experimentally for some time now. 
The thermodynamic state of the strongly interacting medium which is expected to be created in 
a heavy ion collision (HIC) experiment can be specified by four quantities: temperature $T$ and 
three chemical potentials corresponding to the conserved charges baryon number $B$, electric charge 
$Q$ and strangeness $S$, namely $\mu_B$, $\mu_Q$ and $\mu_S$ respectively. The QCD degrees of freedom are sensitive
to these thermodynamic quantities resulting in a rich 
phase diagram~\cite{McLerran:1981pb,Svetitsky:1985ye}. At zero chemical potentials $(\mu_B = \mu_Q = \mu_S = 0)$, first 
principle Lattice QCD (LQCD) Monte-Carlo simulations have shown that QCD undergoes a smooth 
analytic crossover transition from the low $T$ hadron resonance gas (HRG) phase to the high 
$T$ phase of quarks and gluons (QGP)~\cite{Aoki:2006we,Schmidt:2010ss}. At non zero but small 
$\mu_B/T$, recent LQCD results~\cite{Borsanyi:2012cr,Kaczmarek:2011zz} show similar behavior.

However, the QCD phase diagram for $\mu_B/T\gtrsim1$ is far from established. Direct first principle 
techniques of LQCD using Monte-Carlo methods fail due to the sign 
problem~\cite{deForcrand:2010ys,Gupta:2011ma}. This is where QCD like models which have been 
tuned to reproduce LQCD results at zero $\mu_B$ could provide valuable insight about the nature of the QCD medium~\cite{Meisinger:1995ih,
Fukushima:2003fw,Ratti:2005jh,Ghosh:2006qh,Schaefer:2007pw,Mao:2009aq,Schaefer:2008hk,
Lenaghan:2000ey,Fukushima:2010bq,Ray:2011zzb}. Model computations at large $\mu_B/T$ predict 
the possibility of a first order phase transition~\cite{Meisinger:1995ih,Fukushima:2003fw,
Ratti:2005jh,Ghosh:2006qh,Schaefer:2007pw,Mao:2009aq,Schaefer:2008hk,Lenaghan:2000ey,
Fukushima:2010bq,Ray:2011zzb}. Thus the location of the critical point (CP) which is the end 
point of this first order transition line where there is a second order phase transition is 
an important landmark on the QCD phase diagram. Mapping the phase transition line and the CP 
is a major goal of the heavy ion collision experiments~\cite{Mohanty:2013hwa}.

The CP dynamics gives rise to diverging correlation length that result in non-monotonic 
variations of some quantities which have been proposed as plausible observables to 
identify the CP~\cite{Stephanov:1998dy}. For example, moments of conserved charges that 
can be extracted experimentally through event by event analyses are good candidates to hunt 
for the CP~\cite{Stephanov:1999zu,Koch:2005vg}. At zero $\mu_B$, with three flavors of quarks 
there are numerous computations of the susceptibilities both on the 
lattice~\cite{Bernard:2004je,Bernard:2007nm,Cheng:2008zh,Borsanyi:2011sw} as well as in 
models~\cite{Fukushima:2008wg,Fu:2009wy,Fu:2010ay,Bhattacharyya:2010jd,Bhattacharyya:2010ef,
Bhattacharyya:2010ef,Schaefer:2009ui,Wambach:2009ee}. 
Recently, for non-zero but small $\mu_B/T$, susceptibilities have been computed on the 
lattice~\cite{Gupta:2011zzd,Bazavov:2012vg,Borsanyi:2013hza}. Some of these cumulants have also been computed on 
the $\mu_B-T$ plane in models~\cite{Fu:2010ay,Schaefer:2011ex}. 

Such observables are good markers of the CP, as long as they are measured close to the CP, the location of which 
is unknown. Further,
in HIC experiments, the produced fireball has finite size and lifetime which can tame the divergence 
of the correlation length and render it finite. This will blur the effects of singularity in the 
critical region and hence diminish the chances of a direct 
experimental confirmation of the CP~\cite{Berdnikov:1999ph,Nonaka:2004pg,Asakawa:2009aj}. Thus, 
rather than looking at the absolute values of the susceptibilities, sign
structures of the same might be better suited for such studies~\cite{Asakawa:2009aj}. 
Third moments of conserved charges like $B$, $Q$ and energy have been already studied in this 
regard~\cite{Asakawa:2009aj}. These were found to change sign at the hadron-quark phase boundary 
corresponding to peak like structures of second order susceptibilities. Studies based on the 
Polyakov Quark Meson (PQM) model show that higher order cumulants of $B$ and $Q$ become negative 
valued in the transition regime~\cite{Skokov:2011yb,Friman:2011pf}. It has been suggested that such 
distinctive sign structures follow from the scaling functions of the 3-D $O(4)$ universality 
class~\cite{Friman:2011pf}. Higher order generalized susceptibilities for $B$ and their sign 
structure in the phase diagram was studied in the $\l2+1\r$ PQM model~\cite{Schaefer:2011ex}. For all the 
above cases, the negative regions were found very close to the phase boundary and mostly on the 
QGP side. It is interesting to note that LQCD computations along the chemical freeze-out curve as 
determined from HRG analysis of yields show that the kurtosis of $B$ exhibits a change of sign 
around $\sqrt{S_{NN}}\sim20$ GeV~\cite{Gavai:2010zn}. This has been attributed to proximity to 
the CP~\cite{Gavai:2010zn}. Such sign structures of susceptibilities due to the quark-hadron 
transition regime and the CP can be observed if the chemical freeze-out (CFO) curve also passes very close 
to the phase boundary or the dynamics is such that the sign structures are retained during expansion 
between the phase boundary and the CFO curve. However, observation of negative baryonic 
kurtosis has remained elusive so far in the Beam Energy Scan (BES) program at RHIC~\cite{Adamczyk:2013dal}. 
This motivates us to investigate the sign structures of the off-diagonal components of the correlations 
of conserved charges which has so far been ignored. In this paper we work with the $\l2+1\r$ flavor 
PQM model at the mean field level and analyze the sign structures of the cumulants of conserved charges 
on the $\l\mu_B-T\r$ plane with and without CP. We do not find any unique sign structure that could be 
attributed to the presence of the CP alone. On the other hand, there are a few candidates, as summarized 
in Table~\ref{tab.sum}, that are sensitive to the crossover/transition region whether or not there is a CP 
and show a change of sign. Thus these observables are good indicators of the transition regime. Some of 
these off-diagonal susceptibilities were found to exhibit negative regions that extend deep into the hadronic 
side and hence could be more easily accessible to experiments. 

The rest of this paper is organized as follows: In Section \ref{sec.formalism}, we provide the details of the
PQM model and its parameters used in this study. We further define the susceptibilities that are studied 
here and their connections with the corresponding moments of the conserved charges that are experimentally 
measurable. In Section~\ref{sec.results}, we first compare our model computations with those of lattice at 
non-zero but small $\mu_B/T$. We compare, with lattice, some cumulants as well as  the values of the strangeness chemical potential obtained by 
imposing the strangeness neutrality condition. We then present our results on various 
cumulants on the $\mu_B-T$ plane and comment on their usefulness in mapping out the QCD phase diagram, 
namely identifying the transition regime and the location of the CP. Finally, in Section~\ref{sec.conc}, we 
summarize and conclude.

\section{Formalism}
\label{sec.formalism}

The relevant thermodynamic potential $\Omega \l T, \mu_B, \mu_Q, \mu_S \r$ in the $(2+1)$ flavor PQM with 
the inclusion of the vacuum term at a temperature $T$ and chemical potentials $\mu_B$, $\mu_Q$ and 
$\mu_S$ in the mean field approximation is given in Refs.~\cite{Chatterjee:2011jd,Chatterjee:2012np}.
The pressure $Pr$ is given by
\beq
\label{eq.pressure}
Pr\l T, \mu_B, \mu_Q, \mu_S \r=-\Omega \l T, \mu_B, \mu_Q, \mu_S \r\;.
\eeq
The cumulants of the conserved charges are computed by taking appropriate derivatives 
of $Pr$
\beq
   \chi^{BQS}_{ijk}\l T,\mu_B,\mu_Q,\mu_S\r=\frac{\partial^{i+j+k}(Pr/T^4)}
{\partial \l \mu_B/T\r^i\partial \l \mu_Q/T\r^j\partial \l \mu_S/T\r^k}\;.
\label{eq.sus}\eeq
These generalized susceptibilities are related to the moments of the distribution of the conserved
charges such as the mean $M$, variance $\sigma^2$, skewness $S$ etc. The derivatives in eq. (\ref{eq.sus}) 
have been computed numerically using the package {\tt ADOL-C}~\cite{Walther2012Gsw,Wagner:2009pm} which allows efficient 
computation of higher order susceptibilities without further truncation errors.

When relating to heavy ion collision experiments, there are two constraints to 
be met. Since the number of participating nucleons is not fixed a priori, the net baryon number $N_B$ or 
electric charge $N_Q$ can not be fixed independently. But, the ratio of $N_B$ to $N_Q$ can be fixed to 
that of the initial heavy ion used in the experiment ($\sim 2.5$). Since the incoming heavy ions carry 
zero net strangeness $N_S$, the condition of vanishing $N_S$ also has to be imposed. 
\beqa
   N_B/N_Q&=&2.5\label{eq.constr1}\\
   N_S&=&0\label{eq.constr2}
\eeqa
Eqs.~(\ref{eq.constr1}) and (\ref{eq.constr2}) fix $\mu_Q$ and $\mu_S$ respectively. In this work we 
will mainly present our results for 3 different variants of the $\l2+1\r$
flavor PQM model: msig400, msig400-phys and msig600. These differ from each other on the 
choice of the mass of the sigma meson $m_{\sigma}$ and also the way
we treat $\mu_S$ and $\mu_Q$. We have used $m_{\sigma}=400$ MeV for msig400 and msig400-phys while for 
msig600 we put $m_{\sigma}=600$ MeV. While msig400 and msig400-phys possess a CP on the $\l\mu_B-T\r$ 
plane, msig600 has no CP on the phase diagram~\cite{Chatterjee:2011jd}. Thus we are able to study and 
compare the effects of the CP on the sign structures. In order to understand the effects of non-zero 
$\mu_S$ and $\mu_Q$, for msig400 and msig600 we use $\mu_S=0$ and $\mu_Q=0$ while for msig400-phys we 
use $\mu_S$ and $\mu_Q$ as obtained from eqs.~(\ref{eq.constr1}) and (\ref{eq.constr2}). In 
Table~\ref{tab.PQMvar}, we have listed the above scenarios with their descriptions for easy reference
\begin{table}[htb]
\begin{center}
\begin{tabular}{|c|c|c|c|}
\hline
&msig400&msig400-phys&msig600 \\
\hline
\hline
$m_{\sigma}$ (MeV)&400&400&600\\
\hline
$\mu_S$&0&from (\ref{eq.constr1}) and (\ref{eq.constr2})&0\\
\hline
$\mu_Q$&0&from (\ref{eq.constr1}) and (\ref{eq.constr2})&0 \\
\hline
CP&yes&yes&no \\
\hline
\end{tabular}
\end{center}
\caption{The distinct features of the different variants of the PQM model investigated here.}
\label{tab.PQMvar}
\end{table}

\section{Results}
\label{sec.results}

In our earlier works~\cite{Chatterjee:2011jd,Chatterjee:2012np}, we had shown good 
qualitative agreements between LQCD data and PQM model predictions at zero chemical potential 
for a large number of thermodynamic quantities like pressure, entropy density, energy density, 
specific heat, speed of sound etc., and also for some susceptibilities of conserved charges up to 
sixth order. LQCD data are now available also at non-zero but small $\mu_B/T$ and also for 
non-zero $\mu_S$ and $\mu_Q$ satisfying Eqs.~(\ref{eq.constr1}) and (\ref{eq.constr2}
)~\cite{Bazavov:2012vg,Borsanyi:2013hza}. Before we present our model results for the 
susceptibilities on the entire $\mu_B-T$ plane, we perform a comparative study between LQCD 
and our model at small $\mu_B/T$.  Here we will particularly look at the extracted values of 
$\mu_S$ and few ratios of susceptibilities at non-zero $\mu_B/T$.

\subsection{$\displaystyle{\mu_S/\mu_B}$: Model vs Lattice}
\label{subsec.muS}
\begin{figure}
 \includegraphics[scale=0.7]{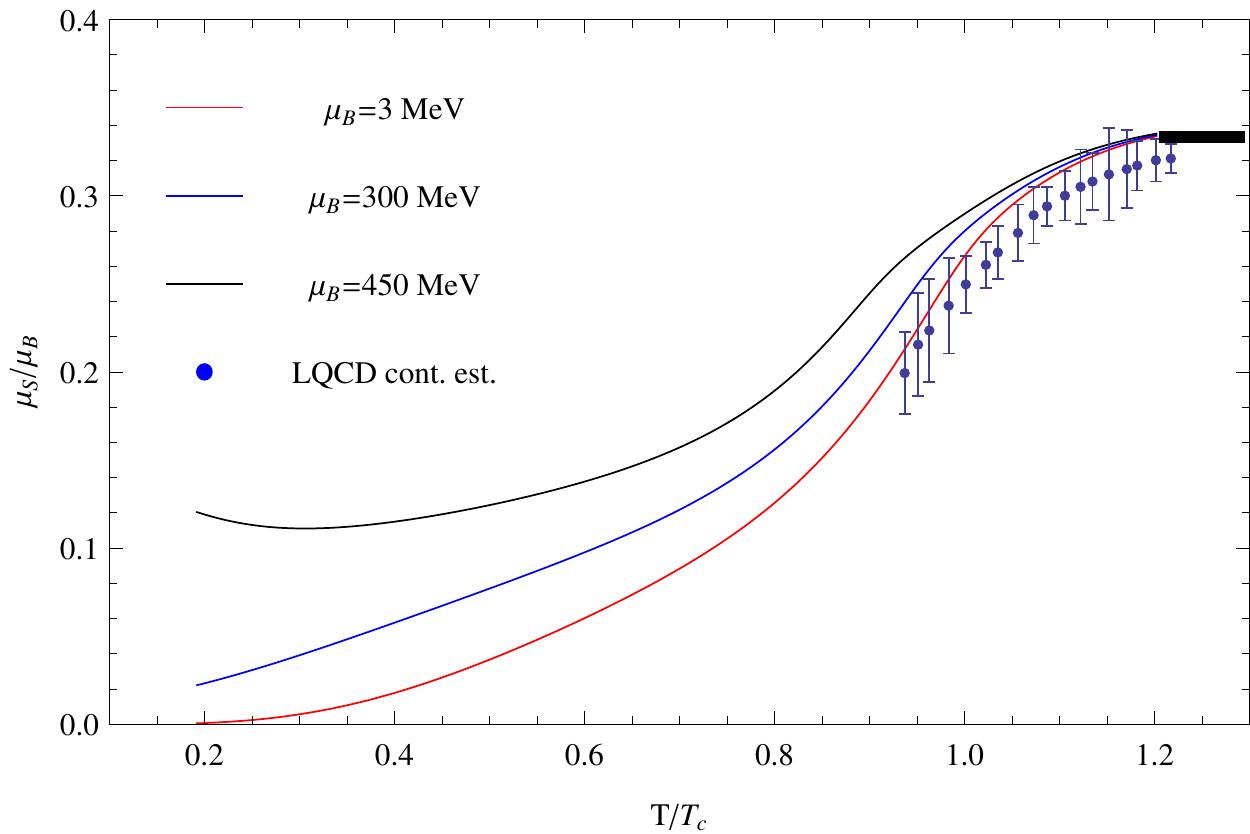}
\caption{Ratio of $\mu_S/\mu_B$ with $T/T_c$ for different values of $\mu_B$ obtained 
in model and compared to leading order results obtained in LQCD~\cite{Bazavov:2014xya}. The thick 
black line denotes the SB limit.}
\label{fig.muslatvspqm}
\end{figure}

For a realistic estimate of thermodynamic observables related to the QGP experiments, it 
is important that the computations are done for non-zero $\mu_S$ and $\mu_Q$ 
obtained from the conditions as imposed by eqs.~(\ref{eq.constr1}) and (\ref{eq.constr2}). We have 
implemented this and extracted the values of $\mu_S$ and $\mu_Q$ for different $T$ and $\mu_B$. 
In Fig.~\ref{fig.muslatvspqm} we show $\mu_S$ normalized to $\mu_B$ for several values of 
$\mu_B=\{3,300,450\}$ MeV. We have compared these model values with the continuum estimates obtained 
in LQCD~\cite{Bazavov:2014xya} and find good agreement. $\mu_S/\mu_B$ seems to have a monotonically 
increasing behavior with $T/T_c$, with a faster rise around $T_c$ and finally saturates 
in the QGP phase to $\sim 1/3$ (shown by the thick black line in Fig.~(\ref{fig.muslatvspqm})). This 
limiting value can be understood easily in the context of a 
Stefan-Boltzmann (SB) gas of ideal quarks and gluons. In the SB limit, only the strange quarks 
which carry both $S$ and $B$ decide the $\mu_S/\mu_B$ value. Now in order to ensure that 
Eq.~(\ref{eq.constr1}) is obeyed, the fugacity factors for strange and anti-strange quarks should be 
unity. Since strange quarks carry $B=1/3$ and $S=-1$ while anti-strange quarks carry $B=-1/3$ and 
$S=1$, it turns out that in order to ensure $B\mu_B+S\mu_S=0$, $\mu_S/\mu_B$ must be $-B/S=1/3$. 
We have ignored $\mu_Q$ in this discussion since its value is much less than both $\mu_S$ and $\mu_B$. 
Thus the SB limit is independent of $\mu_B$ and we see this also in the Fig.~\ref{fig.muslatvspqm} 
where different $\mu_B$ curves all saturate to $1/3$ in the QGP side. On the contrary, in the low 
$T$ regime where the degrees of freedom are hadronic, $\mu_S/\mu_B$ is much more sensitive to 
$\mu_B$. This is because in the hadronic regime unlike the ideal quark gluon gas, strangeness 
carriers can be both: baryonic (eg.$\Lambda$) and non-baryonic or mesonic (eg. kaons). If we 
had a mesonic gas, $\mu_S=0$ will always be the solution of Eq.~\ref{eq.constr1}. Non-zero values of 
$\mu_S$ arise only because of the strange baryons. Since there is a large mass difference of 
the order of $700$ MeV between the lightest strange meson and baryon, depending on $T$ the relative
contribution of the strange baryons differ from that of the strange mesons, rising monotonically with 
$T$. This results in the monotonically increasing behavior of $\mu_S/\mu_B$ with $T/T_c$ for 
constant $\mu_B$. Again, with increasing value of $\mu_B$ the strange baryon contribution increases 
and this results in a larger value of $\mu_S/\mu_B$ for same $T$.

\subsection{$\displaystyle{\frac{\chi^Q_1}{\chi^Q_2},\frac{\chi^B_1}{\chi^B_2}}$: Model vs Lattice}
\label{subsec.rqb}

\begin{figure}
 \includegraphics[scale=0.7]{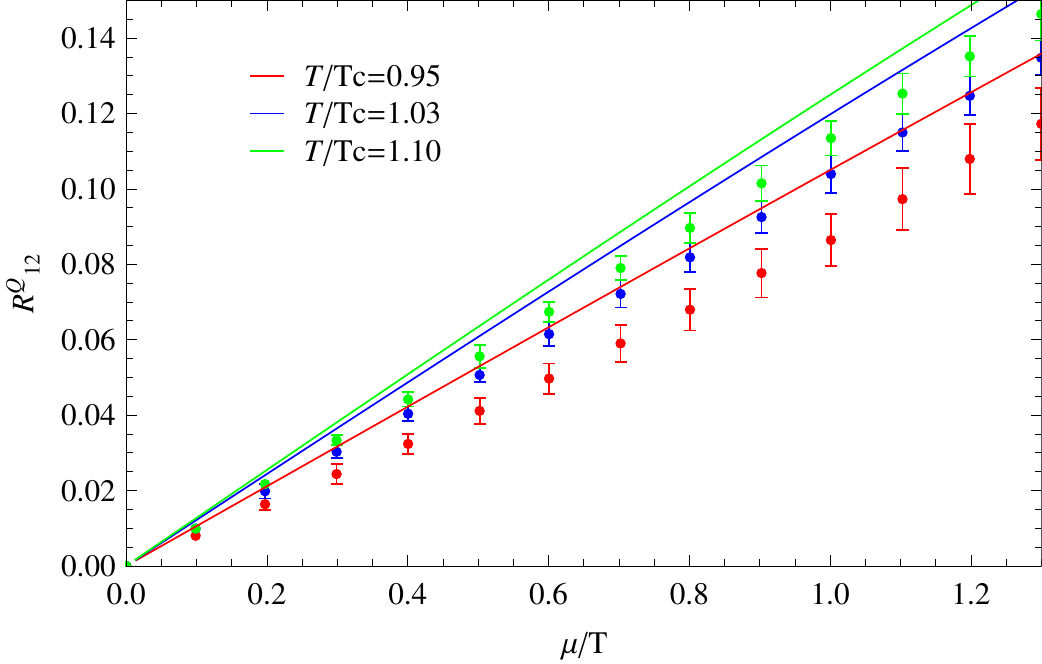}
 \includegraphics[scale=0.7]{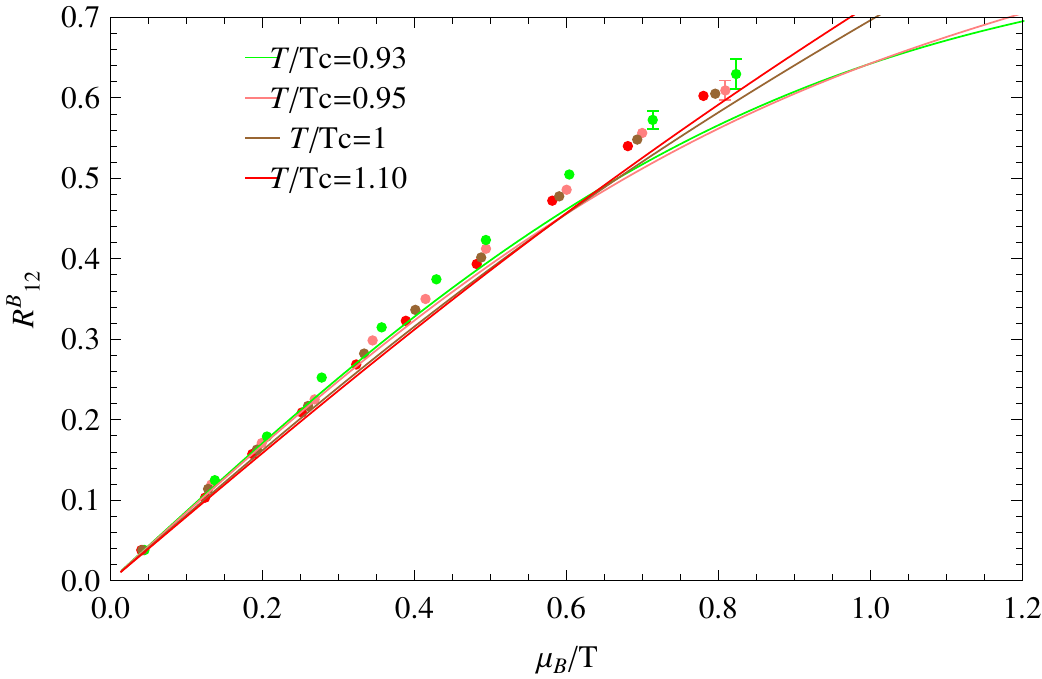}
\caption{Ratios of susceptibilities: $R^Q_{12}=\chi^Q_1/\chi^Q_2$ compared with HotQCD~\cite{Bazavov:2012vg} 
and $R^B_{12}=\chi^B_1/\chi^B_2$ compared with WB~\cite{Borsanyi:2013hza} lattice data.}
\label{fig.latvspqm}
\end{figure}

Recently, a lot of effort has been invested  in computing susceptibilities at
non-zero $\mu_B$ in LQCD in order to confront them with experimentally measured
moments of $B$, $Q$ and $S$~\cite{Gupta:2011zzd,Bazavov:2012vg,Borsanyi:2013hza}. Such a comparison will enable a 
determination of the CFO $T$ and $\mu_B$, bridging the gap between LQCD and experiments. 
This is a complementary program to the already quite successful endeavor of determining 
the freeze-out conditions by comparing the hadron yields between experiments and
hadron resonance gas models~\cite{Cleymans:1998fq,Andronic:2005yp,Becattini:2005xt,Chatterjee:2013yga}. 
In Refs.~\cite{Bazavov:2012vg,Borsanyi:2013hza}, the $\l\mu_B/T\r$ variation of the ratio 
$R^X_{12}=\chi^X_1/\chi^X_2$ for $X=B$, $Q$ have been measured on the lattice. A comparison between 
theory and experiment of these ratios can provide an estimate of $\mu_B$ at CFO~\cite{Bazavov:2012vg}. 
In Fig.~\ref{fig.latvspqm} we have plotted these quantities
as obtained in PQM and compared them with the HotQCD lattice data~\cite{Bazavov:2012vg} as well as WB 
lattice data~\cite{Borsanyi:2013hza}. Both lattice as well as PQM
show an almost linear variation of $R^Q_{12}$ with $\mu_B/T$ in the range $0<\mu_B/T<1$.
The model over predicts the lattice data in the entire range. For example, at $\mu_B/T\sim1$, the 
model is about $20\%$ more than the lattice. We should note that there is an uncertainty in the
determination of $T_c$ adjusting which it is possible to get a better agreement between model and
lattice. However, here our main intention in performing this comparison is to demonstrate the good
qualitative agreement between LQCD and PQM model predictions. This gives us faith to trust the PQM results on the $\mu_B-T$ plane 
where there is yet to be any lattice data. Thus, having set the platform for a discussion of the
PQM results at non-zero $\mu_B$, we go into the next section where we report 
on the  novel qualitative features of the sign structures of the generalized susceptibilities
whose measurements in heavy ion collision experiments could shed light on the features of the 
QCD phase diagram.

\subsection{Sign Structures of Susceptibilities: Model}
\label{subsec.sus}

Having found good qualitative agreement between PQM and LQCD at non-zero $\mu_B$,
we will now analyze the sign structures of various fluctuations and correlations of
conserved charges on the $\l\mu_B-T\r$ plane. We shall focus mainly on the transition
regime between the HRG and QGP phases. The transition region can be broadly classified 
into three categories: firstly for small $\mu_B/T$ there is a smooth crossover and no 
true phase transition. Secondly, for large enough $\mu_B/T$, we expect a first order phase 
transition. Thirdly, in the intermediate range of $\mu_B/T$, one thus expects a critical 
region with a second order phase transition at the CP where the line of first 
order phase transition meets the crossover line. The existence and location of this CP is 
a topic of intense current research. In PQM, the location of the CP is highly sensitive to 
the value of $m_{\sigma}$ used. For example, with $m_{\sigma} \ge 600$ MeV one finds no 
CP~\cite{Chatterjee:2011jd}. In order to understand the distinguishing features of the
QCD phase diagram with and without a CP and also the effect of non-zero $\mu_Q$ and $\mu_S$, 
we have computed the susceptibilities for three different variants of PQM as outlined in 
Table~\ref{tab.PQMvar}: msig400, msig400-phys and msig600. We will now report our findings 
on the sign structure of several susceptibilities up to sixth order in these different regions 
of the QCD phase diagram for the three cases.

Before we proceed, please note that all figures shown henceforth will be plots on the $\mu_B - T$ plane. 
In fact, since we are interested only in the qualitative features of negative correlations, we scale the 
$T$ axis ($x$-axis) by $T_c$ (the cross-over temperature at $\mu_B=0$) and the $\mu_B$ axis (y-axis) by $\mu_c$ (the $\mu_B$ 
value where the first order phase transition line meets the $\mu_B$ axis). In all plots we show the smooth 
cross-over curve in blue dashed lines. The CP is denoted by a red dot whereas a black (thick) solid line 
denotes the first order phase transition line. Contours of various susceptibilities are shown with boxed 
numbers indicating the numerical value of the susceptibilities along those contours. Finally, regions of 
negative correlation are indicated by orange (shaded) regions.

\subsubsection{Diagonal Susceptibilities up to 4th order}
\label{subsubsec.diag}
\begin{figure}
\vspace{-1.5cm}
\includegraphics[scale=0.8]{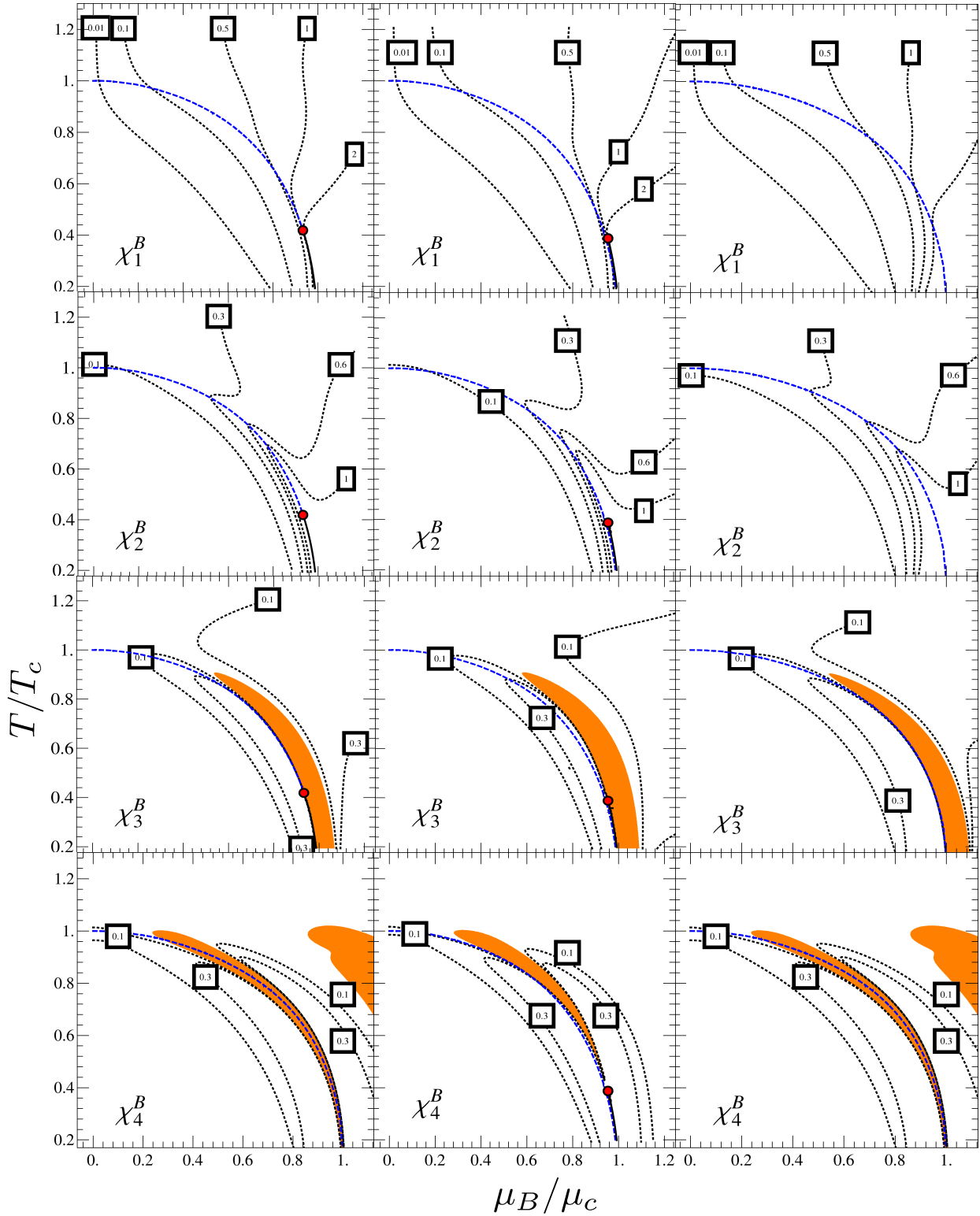}
\caption{Contours (black dotted) for diagonal susceptibilities up to fourth order, shown on the $\mu_B -T $ plane. Also shown are the cross-over curve (blue dashed) the CP (red dot) and the first order transition curve (black solid). Boxed number indicate the value of the susceptibilities along that contour. Negative regions are shown in orange (shaded). The three columns are for the three different variants of the PQM model in Table~\ref{tab.PQMvar}. Column 1: msig400 , Column 2 : msig400-phys and Column 3: msig600}
\label{fig.diagB}\end{figure}

\begin{figure}
\includegraphics[scale=0.8]{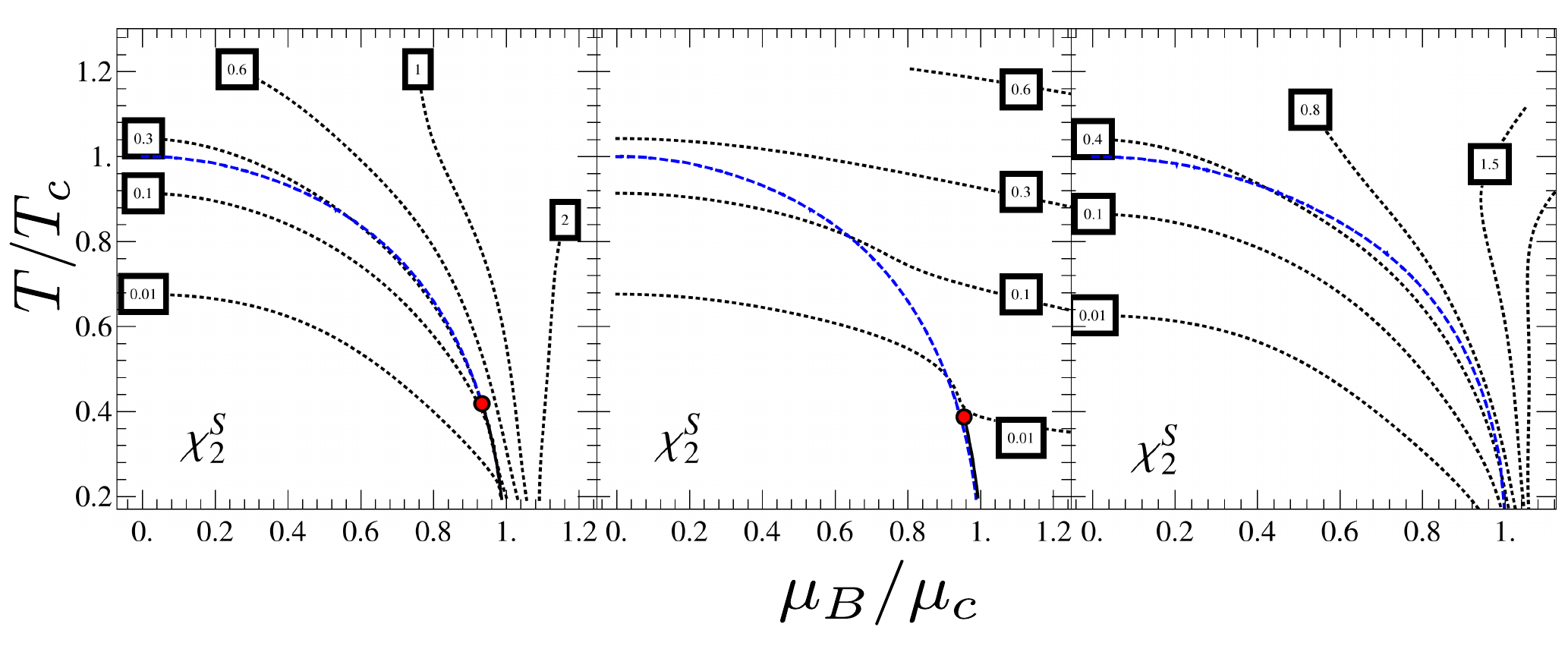}
\caption{Contours (black dotted) for $\chi^{S}_{2}$ shown on the $\mu_B -T $ plane. Also shown are the cross-over curve (blue dashed) the CP (red dot) and the first order transition curve (black solid). Boxed number indicate the value of the susceptibilities along that contour. Negative regions are shown in orange (shaded). The three columns are for the three different variants of the PQM model in Table~\ref{tab.PQMvar}. Column 1: msig400 , Column 2 : msig400-phys and Column 3: msig600}
\label{fig.diags}
\end{figure}

We have shown the baryonic susceptibilities up to 4th order in Figure.~\ref{fig.diagB}.
Experimentally only net proton number is obtained as the neutrons are
never observed. Cumulants of net proton number act as proxy to the susceptibilities of 
the net baryon number. Up to 2nd order there are no zero contours and therefore no regions of negative correlation. The first appearance
of a negative region is for $\chi^B_3$. It spans through the crossover as well
as first order phase transition regions. For all the three cases studied here, 
the negative region is located just above the hadron-quark transition on the 
QGP side. For $\chi^B_4$, the negative region is only in the 
crossover region and terminates at the CP. Negative values for 
$\chi^B_4/\chi^B_2$ have also been measured on lattice at non-zero 
$\mu_B$~\cite{Gavai:2010zn}. For the case of msig600,
even though the CP is absent, there is a negative region in the 
transition regime that stretches  all the way up to very small T. Thus, we see
that negative regions in $\chi^B_3$ and $\chi^B_4$ only imply the proximity of
the hadron-quark transition region that may or may not include a CP. Thus negative 
values of $\chi^B_3$ and $\chi^B_4$ do not necessarily imply the existence of a CP. 
We have not shown plots of the susceptibilities of $Q$ as the sign structure for the 
electric charge susceptibilities look similar to that of the baryonic ones. The
contours of $S$ susceptibilities on the $\l\mu_B-T\r$ plane are devoid of any interesting 
sign structures. The $\chi^S_2$ susceptibilities are shown in Fig.~\ref{fig.diags}. We 
note that in msig400-phys where the physical conditions of eqs. (\ref{eq.constr1}) and (\ref{eq.constr2}) are 
imposed, the diagonal strange susceptibilities are affected strongly and are distinctly 
different from what one obtains in msig400 and msig600. In msig400-phys the variation of 
$\chi^S_2$ along $\mu_B$ becomes gentler with the contours being almost parallel to the 
$\mu_B$ axis.

\subsubsection{Mixed susceptibilities of order two}

\begin{figure}
\includegraphics[scale=0.8]{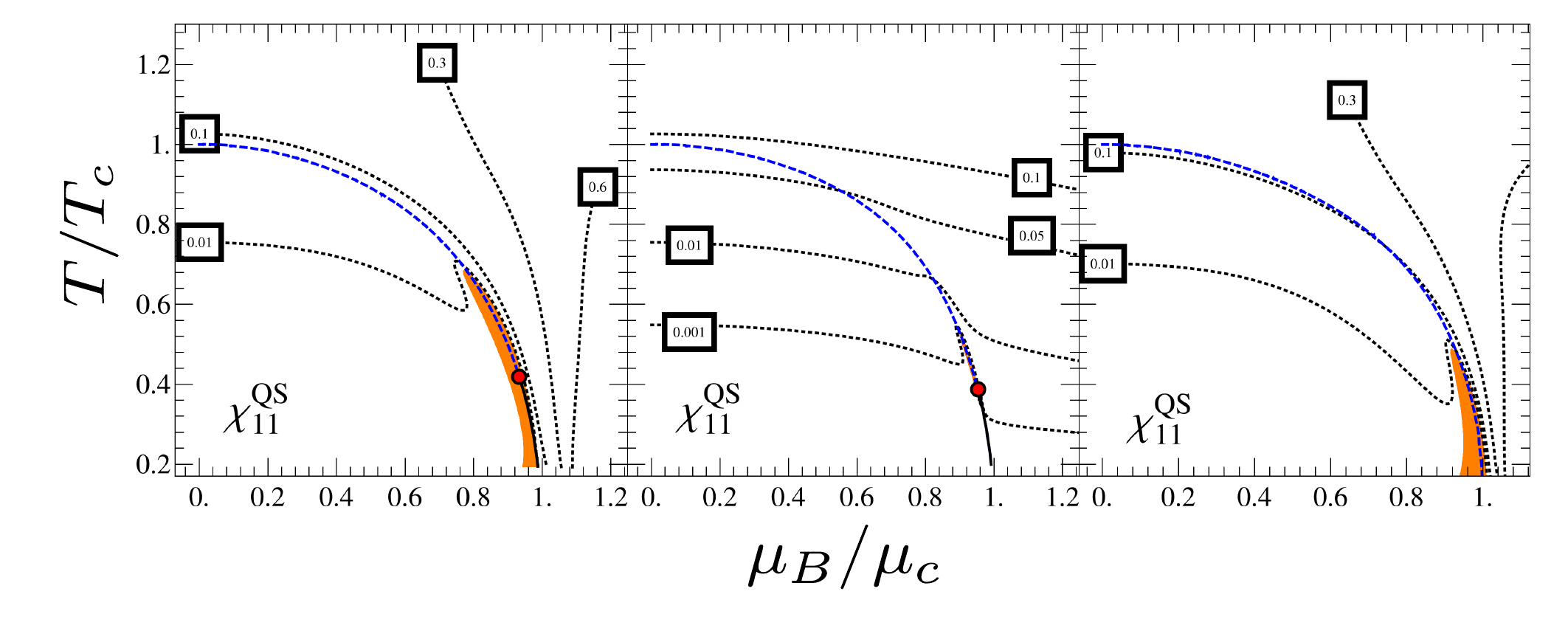}
\caption{Contours (black dotted) for $\chi^{QS}_{11}$, shown on the $\mu_B -T $ plane. Also shown are the cross-over curve (blue dashed) the CP (red dot) and the first order transition curve (black solid). Boxed number indicate the value of the susceptibilities along that contour. Negative regions are shown in orange (shaded). The three columns are for the three different variants of the PQM model in Table~\ref{tab.PQMvar}. Column 1: msig400 , Column 2 : msig400-phys and Column 3: msig600}
\label{fig:mix2}
\end{figure}

Neither $\chi^{BQ}_{11}$ nor $-\chi^{BS}_{11}$ have any interesting sign structure. Neither have any negative regions. On the other hand,
 $\chi^{QS}_{11}$ has a  narrow region where it turns negative for 
msig400 and msig600. Interestingly, this region of negative susceptibilities 
lies in the hadronic side.
Fig.~\ref{fig:mix2} shows $\chi^{QS}_{11}$ which has negative values for large $\mu_B$ 
close  to the transition region and within the hadronic phase. For msig400 we see that the region is 
close to the CP. For msig400-phys we observe that regions of negative correlation are in fact reduced and very 
close to the CP. Since negative correlations are also seen for msig600, we infer that 
a measurement of negative values of this correlation does not necessarily confirm a CP. 

\subsubsection{Mixed susceptibilities of order three}
\begin{figure}
\includegraphics[scale=0.8]{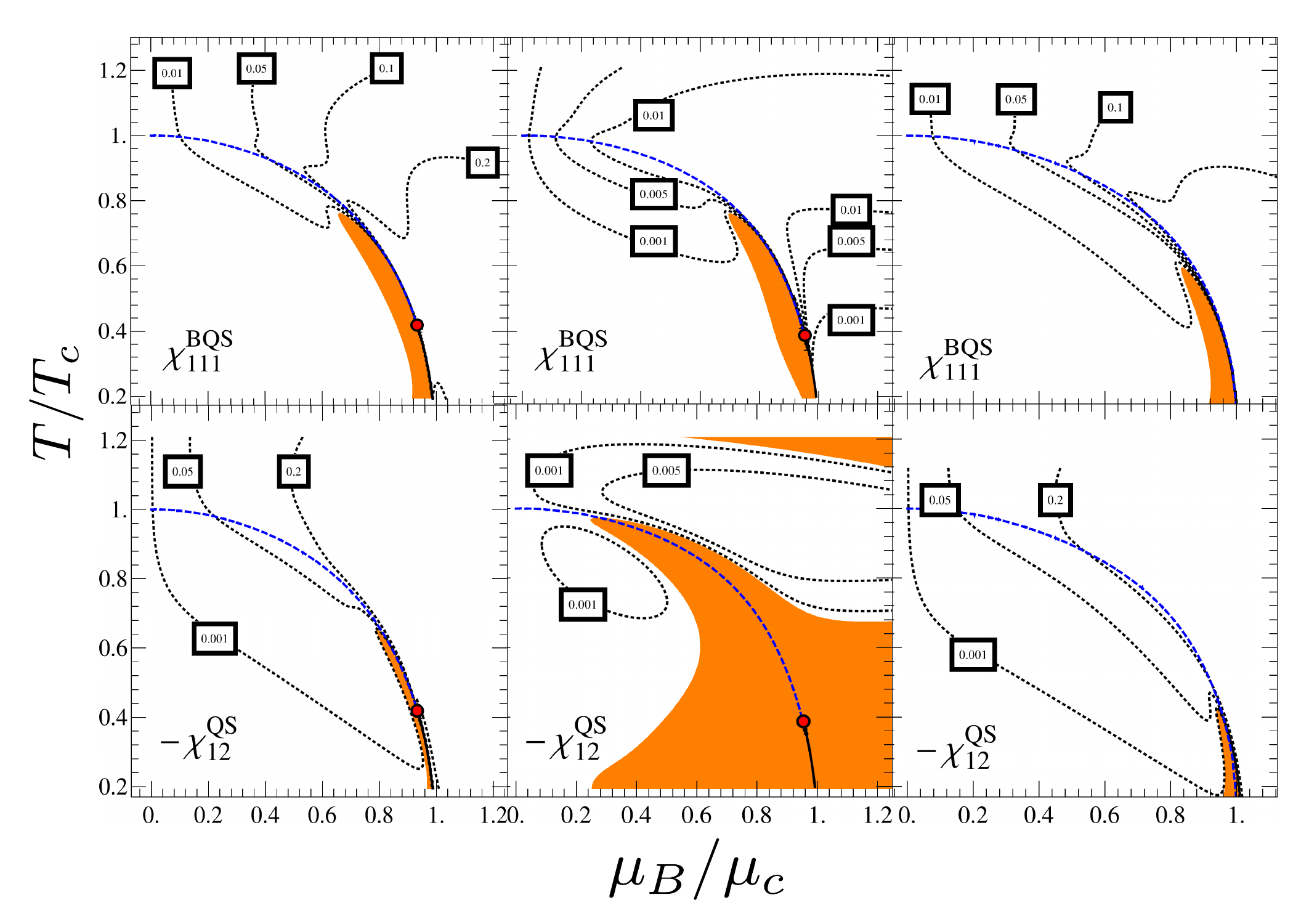}
\caption{ Contours (black dotted) for $\chi^{BQS}_{111}$ and $-\chi^{QS}_{12}$, shown on the $\mu_B -T $ plane. Also shown are the cross-over curve (blue dashed) the CP (red dot) and the first order transition curve (black solid). Boxed number indicate the value of the susceptibilities along that contour. Negative regions are shown in orange (shaded). The three columns are for the three different variants of the PQM model in Table~\ref{tab.PQMvar}. Column 1: msig400 , Column 2 : msig400-phys and Column 3: msig600}
\label{fig:mix3}
\end{figure}

The third order mixed susceptibilities show more interesting behavior. In Fig.\ref{fig:mix3}, 
we show $\chi_{12}^{QS}$ and $\chi_{111}^{BQS}$. For msig400-phys, there is a large region deep 
into the HRG phase where $\chi_{12}^{QS}$ is negative. Thus, there is a possibility that the 
CFO curve passes through this region and these negative regions could be observed in experiments. 
Interestingly, for both msig400 and msig600 where $\mu_Q$ and $\mu_S$ are zero this negative 
region is highly localized on the first order line. $\chi^{BS}_{12}$ which is not shown here has 
small negative valued region along the crossover line from the CP in msig400. However, this effect 
is absent in msig400-phys and msig600 and therefore probably cannot be used to probe the CP 
in experiment.  For $\chi_{111}^{BQS}$, all the variants show negative region in the
hadronic side, though they do not extend as deep as in the case of $\chi_{12}^{QS}$. Interestingly, 
the region of negative correlation extends deeper into the hadronic region for msig400-phys For
the remaining off diagonal susceptibilities of order 3 we find that none of them have regions of 
negative correlation in the hadronic phase, although some of them have negative regions in the 
QGP phase close to the transition curve.

\subsubsection{Mixed susceptibilities of higher order}
\begin{figure}
\vspace{-1.5cm}
\includegraphics[scale=0.8]{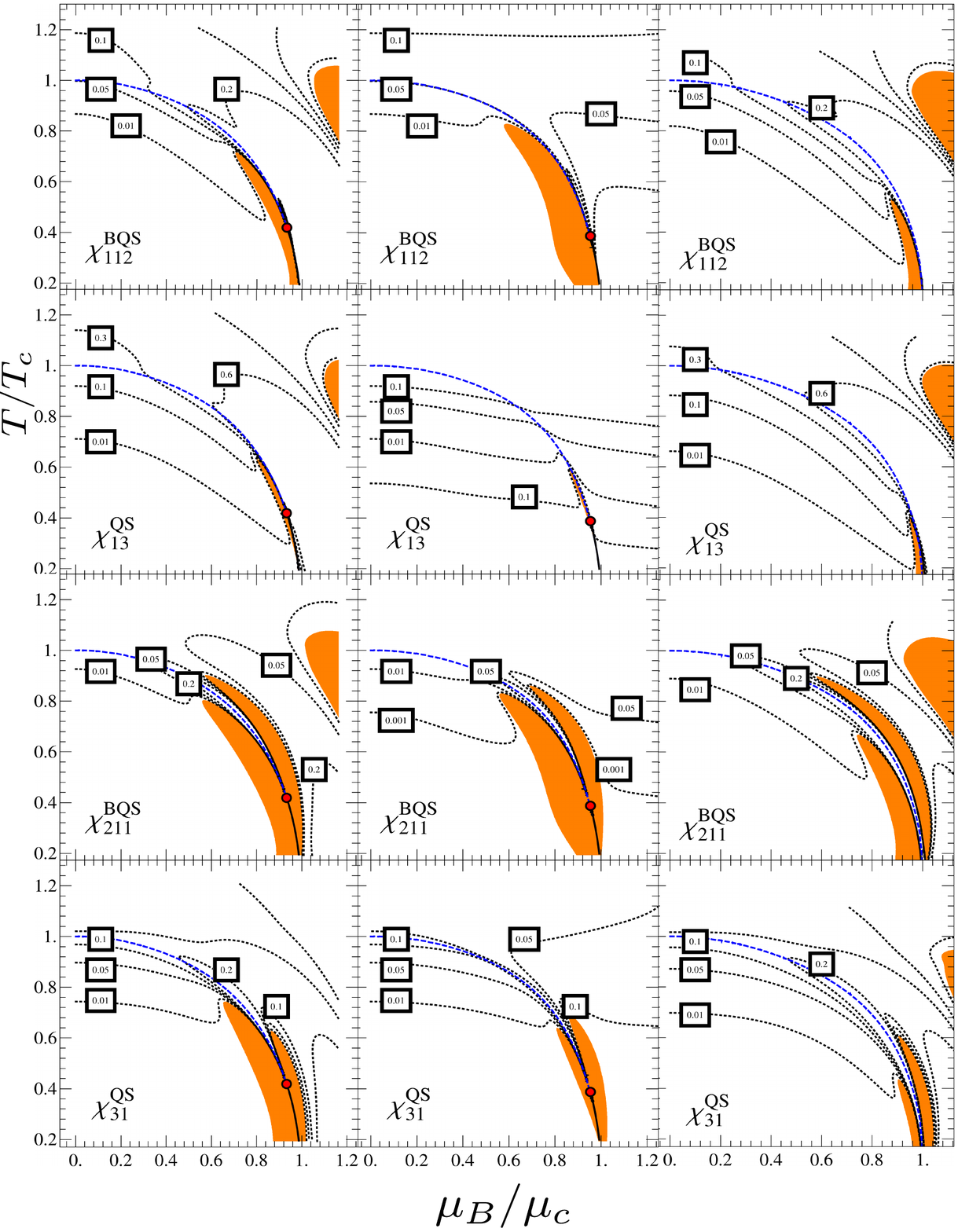}
\caption{ Contours (black dotted) for fourth order off-diagonal susceptibilities, shown on the $\mu_B -T $ plane. Also shown are the cross-over curve (blue dashed) the CP (red dot) and the first order transition curve (black solid). Boxed number indicate the value of the susceptibilities along that contour. Negative regions are shown in orange (shaded). The three columns are for the three different variants of the PQM model in Table~\ref{tab.PQMvar}. Column 1: msig400 , Column 2 : msig400-phys and Column 3: msig600}
\label{fig:mix4}
\end{figure}

\begin{figure}
\includegraphics[scale=0.8]{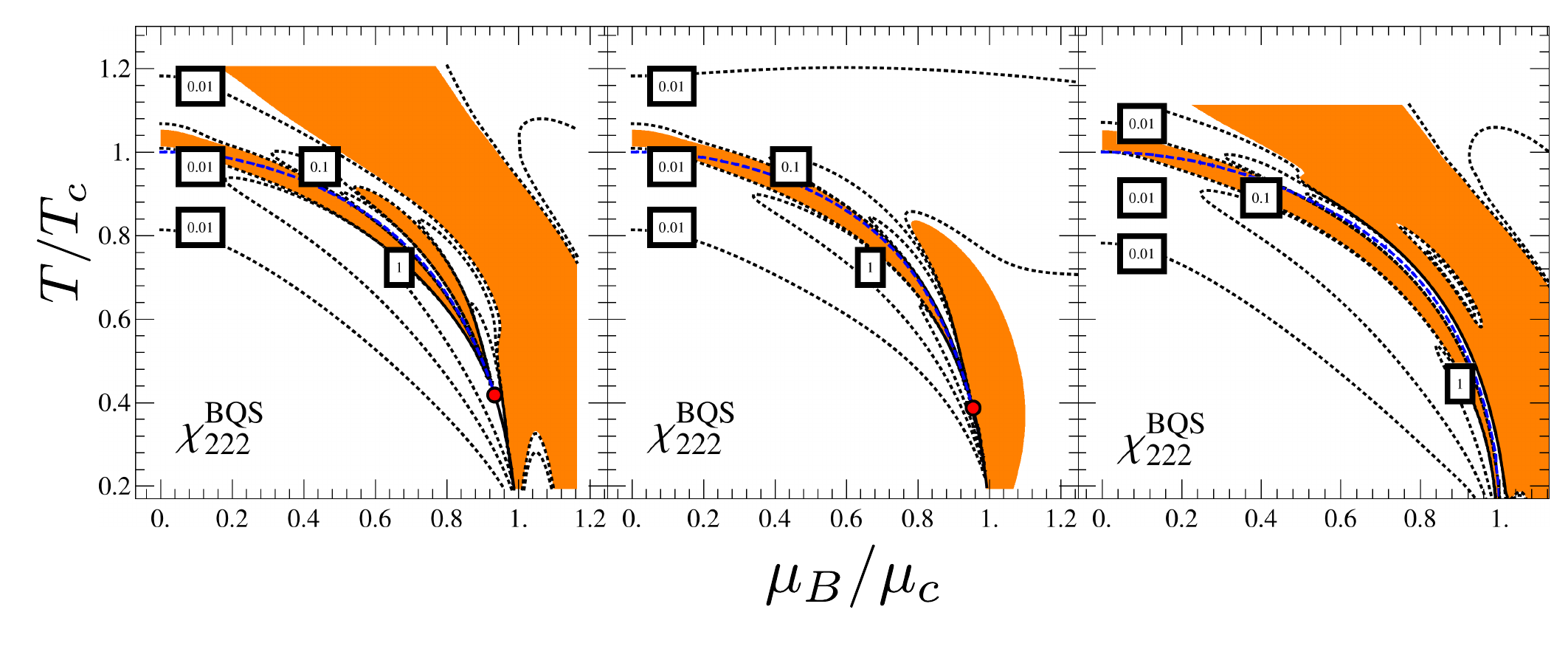}
\caption{Contours (black dotted) $\chi^{BQS}_{222}$, shown on the $\mu_B -T $ plane. Also shown are the cross-over curve (blue dashed) the CP (red dot) and the first order transition curve (black solid). Boxed number indicate the value of the susceptibilities along that contour. Negative regions are shown in orange (shaded). The three columns are for the three different variants of the PQM model in Table~\ref{tab.PQMvar}. Column 1: msig400 , Column 2 : msig400-phys and Column 3: msig600}
\label{fig:mix6}
\end{figure}

In Fig.~\ref{fig:mix4} we show only those fourth order off-diagonal susceptibilities that have  
regions of negative correlation that penetrate the hadronic phase significantly. $\chi_{112}^{BQS}$ 
and $\chi_{13}^{QS}$ have negative values only in the hadronic phase
\footnote{Other negative regions seen in the plot are deep in the QGP phase and we disregard them.}. We observe that while  
$\chi_{112}^{BQS}$ extends deep into the hadronic phase, $\chi_{13}^{QS}$  is highly localized near 
the transition region and the CP. Also while the negative region for $\chi_{112}^{BQS}$ appears 
extended in msig400-phys as compared to the other cases, the opposite effect 
is observed in $\chi_{13}^{QS}$ where the negative region diminishes slightly.
For $\chi^{BQS}_{211}$ and $\chi^{QS}_{31}$ regions of negative correlation exist on either side of 
the QCD transition line at large $\mu_B$ and close to the CP. These regions appear as two separate 
lobes that merge across the first order transition curve and separate above the CP where the nature 
of the transition is that of a cross-over. $\chi^{BQS}_{211}$ shows a slightly enhanced region of 
negative correlation for the physical conditions of msig400-phys in comparison to the other cases while 
the negative regions are reduced for $\chi^{QS}_{31}$ in msig400-phys. In Fig.\ref{fig:mix6} 
we show the contours for $\chi^{BQS}_{222}$ which has a behavior similar to 
the diagonal susceptibilities, with the region of negative correlation closely following the QCD 
transition curve and proceed into the QGP side.

\subsection{Tracking the QCD phase transition}

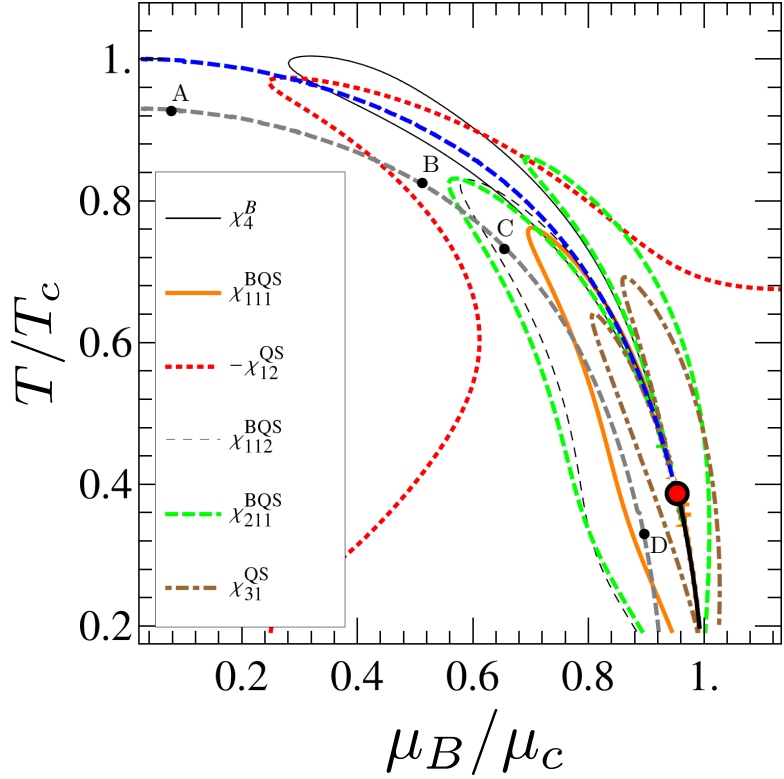
\begin{figure}
\includegraphics[scale=0.8]{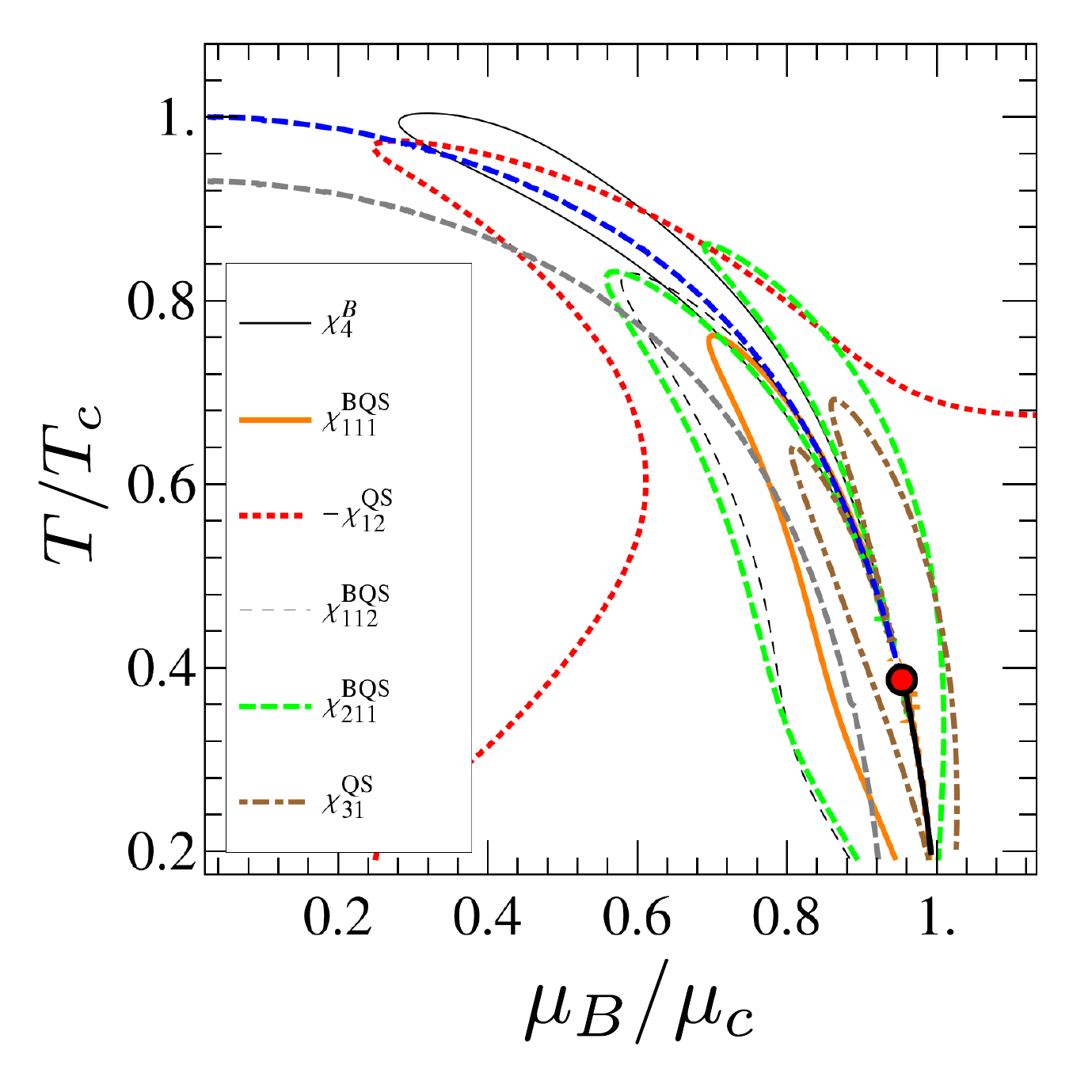}
\begin{picture}(0,0)
\put(-250,269){A}
\put(-250,265){\circle*{4}}
\put(-155,242){B}
\put(-155,238){\circle*{4}}
\put(-127,218){C}
\put(-124,213){\circle*{4}}
\put(-69,98){D}
\put(-71,105){\circle*{4}}
\end{picture}

\caption{Contours of negative regions for various susceptibilities for msig400-phys.
The gray thick dashed curve corresponds to the proxy freeze-out line.
Also shown are the cross-over curve (blue dashed) the CP (red dot) and the first order transition curve (black solid).}
\label{fig:sumplot}
\end{figure}


\begin{table}
	\centering
	\begin{tabular}{|c|p{1.7cm}|p{1.7cm}|p{2 cm}|p{1.5cm}|}
		\hline
		Susceptibility      & Negative Region& Negative in Hadronic Phase& Range & R \\ 
		\thickhline
		$\chi^{B}_{1}$      & \xmark    & --   &  --   &  --    \\ \hline
		$\chi^{B}_{2}$      & \xmark    & --   &  --   &  --   \\ \hline
		\rowcolor{Gray}
		$\chi^{B}_{3}$      & \cmark    & \xmark   & (0.6,0.9)-E    &  --   \\ \hline
		\rowcolor{lmv}
		$\chi^{B}_{4}$      & \cmark    & \cmark   &(0.3,1.0)-CP     & 0.98     \\ \hline
		
		\thickhline
		$\chi^{Q}_{1}$      & \xmark    & --   &  --   &  --   \\ \hline
		$\chi^{Q}_{2}$      & \xmark    & --   &  --   &  --   \\ \hline
		\rowcolor{Gray}
		$\chi^{Q}_{3}$      & \cmark    & \xmark   & (0.9,0.5)-E    & --    \\ \hline
		$\chi^{Q}_{4}$      & \cmark    & \xmark   &(0.7,0.9)-CP     & --    \\ 
		
		\thickhline
		$\chi^{S}_{2}$      & \xmark    & --   &  --   &  --   \\ \hline
		$\chi^{S}_{3}$      & \xmark    & --   &  --   &  --   \\ \hline
		$\chi^{S}_{4}$      & \xmark    & --   &  --   &  --   \\ 
		\thickhline
		$\chi^{BQ}_{11}$    & \xmark    & --   &  --   &  --   \\ \hline
		$\chi^{BS}_{11}$    & \xmark    & --   &  --   &  --   \\ \hline
		\rowcolor{lmv}
		$\chi^{QS}_{11}$    & \cmark    & \cmark   &  (0.9,0.6)-CP   &  0.99   \\ 
		
		\thickhline
		\rowcolor{lmv}
		$\chi^{BQS}_{111}$  & \cmark    & \cmark  & (0.7,0.7)-E   & 0.91    \\ \hline
		\rowcolor{Gray}
		$\chi^{BQ}_{12}$    & \cmark    & \xmark   & (0.9,0.7)-E   &  --   \\ \hline
		\rowcolor{Gray}
		$\chi^{BQ}_{21}$    & \cmark    & \xmark   & (0.5,0.9)-E   & --    \\ \hline
		$\chi^{BS}_{12}$    & \xmark    & --   &  --   &  --   \\ \hline
		\rowcolor{Gray}
		$\chi^{BS}_{21}$    & \cmark    & \xmark   & (0.9,0.6)-CP   & --    \\ \hline
		\rowcolor{lmv}
		$\chi^{QS}_{12}$    & \cmark    & \cmark   &    &     \\ \hline
		\rowcolor{Gray}
		$\chi^{QS}_{21}$    & \cmark    & \xmark   &(0.6,0.9)-E    & --    \\ 
		
		\thickhline
		\rowcolor{lmv}
		$\chi^{BQS}_{211}$  & \cmark    & \cmark   &(0.6,0.8)-E    & 0.84    \\ \hline
		\rowcolor{lmv}
		$\chi^{BQS}_{121}$  & \cmark    & \cmark   & (0.7,0.9)-CP   & 0.99    \\ \hline
		\rowcolor{lmv}
		$\chi^{BQS}_{112}$  & \cmark    & \cmark   & (0.6,0.8)-E   & 0.84    \\ \hline
		\rowcolor{lmv}
		$\chi^{BQ}_{22}$    & \cmark    & \cmark   & (0.4,0.9)-CP   & 0.99     \\ \hline
		\rowcolor{Gray}
		$\chi^{BS}_{22}$    & \cmark    & \xmark   &(0.6,0.9)-E    & --    \\ \hline
		\rowcolor{Gray}
		$\chi^{QS}_{22}$    & \cmark    & \xmark   & (0.9,0.7)-E   &  --   \\ \hline
		\rowcolor{lmv}
		$\chi^{BQ}_{31}$    & \cmark    & \cmark   & (0.2,0.1)-CP    & 0.98     \\ \hline
		\rowcolor{lmv}
		$\chi^{BQ}_{13}$    & \cmark    & \cmark   &(0.4,1.0)-CP    & 0.99    \\ \hline
		\rowcolor{lmv}
		$\chi^{QS}_{31}$    & \cmark    & \cmark   & (0.8,0.6)-E   & 0.96    \\ \hline
		\rowcolor{lmv}
		$\chi^{QS}_{13}$    & \cmark    & \cmark   & (0.8,0.6)-CP   & 0.99    \\ \hline
		\rowcolor{lmv}
		$\chi^{BS}_{31}$    & \cmark    & \cmark   & (0.4,1.0)-CP   & 0.99    \\ \hline
		$\chi^{BS}_{13}$    & \xmark    & --   &  --   &  --   \\ 
		
		\thickhline
		\rowcolor{Gray}
		$\chi^{BS}_{23}$    & \cmark    & \xmark   & (0.4,0.9)-CP   & --     \\ \hline
		\rowcolor{lmv}
		$\chi^{BQS}_{113}$  & \cmark    & \cmark   & (0.8,0.7)-E   & 0.92    \\ \hline
		$\chi^{BS}_{14}$    & \xmark    & --   &  --   &  --   \\ 
		
		\thickhline
		\rowcolor{lmv}
		$\chi^{BQS}_{222}$  & \cmark    & \cmark   & ()-CP   & 0.96    \\ \hline
		\rowcolor{lmv}
		$\chi^{BS}_{33}$    & \cmark    & \cmark   &(0.2,1.0)-CP    & 0.97    \\ \hline
		\rowcolor{lmv}
		$\chi^{BQS}_{213}$  & \cmark    & \cmark   & (0.6,0.8)-E   & 0.87     \\ \hline
		\rowcolor{lmv}
		$\chi^{BQS}_{123}$  & \cmark    & \cmark   & (0.5,0.9)-CP   & 0.98    \\ \hline
		\rowcolor{lmv}
		$\chi^{BS}_{24}$    & \cmark    & \cmark   & (0.3,1)-E   &  0.99   \\ \hline
		\rowcolor{lmv}
		$\chi^{BQS}_{114}$  & \cmark    & \cmark   & (0.7,0.7)-E   & 0.87    \\ \hline
		$\chi^{BS}_{15}$    & \xmark    & --   &  --   &  --   \\ \hline
		
	\end{tabular}
	\caption{Summary Table: A list of various susceptibilities with a description of their negative regions. The second column indicates the presence of 
		regions of negative correlation with a tick and absence with a cross. The third column indicates whether any part of 
		the negative region lies in the hadronic phase. 
		The fourth column corresponds 
		to the approximate range of the region (see text for meaning). The last column denotes the extent 
		of the negative region into the hadronic side (see text for details). Here R corresponds to a scaling factor as described in the text. }
	\label{tab.sum}
\end{table}

After a brief description of the essential features of the sign structure 
of various susceptibilities in the discussion above, we would like to indicate how a measurement of the same might be 
useful in locating the QCD phase transition curve as well as the CP.
In Fig.~\ref{fig:sumplot}, we show along with the phase transition curve, negative regions of a few select susceptibilities, 
namely $\chi^{B}_{4}$, 
$\chi^{BQS}_{111}$, $-\chi^{QS}_{12}$, $\chi^{BQS}_{112}$, $\chi^{BQS}_{211}$ and $\chi^{QS}_{31}$.
We include the diagonal $\chi^{B}_{4}$ susceptibility in this plot to accentuate the fact the off-diagonal susceptibilities 
have regions of negative correlation significantly deeper in the hadronic phase.
It is our proposal that a systematic measurement of these off-diagonal susceptibilities can provide valuable information 
about the proximity of the 
phase transition curve as well as the CP.

In order to illustrate this point, let us consider a hypothetical CFO curve as indicated by the (gray dashed line). 
The curve has been drawn by scaling the phase transition curve down by a factor $R=0.93$ and hence our CFO curve follows 
the phase transition curve. We have marked four points (labeled: A, B, C, and D) at various values of $(\mu_B,T)$ along 
this curve that indicate possible freeze-out points at which the sign of the susceptibilities can be determined.
Point A, at high T and low $\mu_B$ (akin to the experiments at LHC) would not show negative values of the susceptibilities 
under consideration. At lower energies and larger $\mu_B$, we approach point B where we might obtain the negative values 
for one of the susceptibilities, namely $-\chi^{QS}_{12}$. As we proceed to point C along the CFO, we now have three negative 
susceptibilities; $-\chi^{QS}_{12}$, $\chi^{BQS}_{112}$ and $\chi^{BQS}_{111}$. Finally at point D, in addition to these 
three susceptibilities, $\chi^{BQS}_{111}$ is also negative. This indicates that point D is closest to the CP. 

One must be wary that the CFO curve might have a more complicated shape than the simple scaled curve that we have used. In which case 
the CFO curve may miss some of the negative regions. Nevertheless, measurement of negative values of the susceptibilities 
we have considered here are indicative of proximity to the phase transition region. Most importantly, since these 
susceptibilities are negative deep in the hadronic phase, the CFO curve is more likely to pass through them.

Before we conclude, we summarize our findings in Table~\ref{tab.sum}. In this table we list various susceptibilities and describe some of their properties. In the second column we of the table we indicate the presence of region of negative correlation by a tick mark and the absence of one by a cross. In the third column we indicate whether there there is a region of negative correlation in the hadronic phase. In the last two columns we map out regions of negative correlation. 
This we do by first observing that a common characteristic of negative regions is that they originate at some point of the $\mu_B -T$ plane and follow the phase transition curve. The negative regions terminate either at the CP or follow the phase transition curve all the way up to the end (E) at zero T.
The fourth column thus indicates the point at which the negative region begins $(\mu_B/\mu_c,T/T_c)$ and whether it follows the phase transition curve up to the CP or to the end (E).
Finally, for those susceptibilities that have negative regions in the hadronic phase we give the least value of the scale factor (R) for which our hypothetical CFO curve (described above) just touches the negative region.
As can be gleaned from the table, there are several useful candidates that could be used to locate the QCD phase transition curve and CP.

\section{Conclusion}
\label{sec.conc}

The sign structure of diagonal susceptibilities have been studied previously with negative 
regions found in the HRG-QGP transition regime and mostly in the QGP side. Thus only if CFO 
occurs close to the phase transition regime are these negative regions likely to be 
observed. HRG analysis of yields reveal that the CFO curve is close to the QCD 
transition curve for small $\mu_B$ as in the case of LHC~\cite{Cleymans:1998fq,Andronic:2005yp,
Becattini:2005xt,Chatterjee:2013yga}. At SPS and future experiments like FAIR where $\mu_B$ is 
large, the CFO curve could be 
further away from the transition curve and therefore it is likely that these experiments may 
miss the negative regions. In this study we find that for particular choices of off-diagonal 
susceptibilities as listed in Table~\ref{tab.sum}, negative regions could extend deep into the hadronic 
phase and should be accessible to the experiments even at large values of $\mu_B$. Our study suggest 
that the correlations of conserved charges exhibit a rich sign structure in the $\mu_B-T$ plane that 
can be accessed by experiments. These can guide us in our search for the CP as well as the 
hadron-quark phase transition line. Their detection will also signal unambiguously the creation of 
a  novel phase of QCD matter unlike the HRG phase. However, we should caution the reader that unlike 
critical exponents, sign structures of cumulants and the region over which they are negative are 
model dependent. Thus, our results should be viewed
with caution when we want to extrapolate to real QCD. Another caveat is that , currently this study is at the mean field 
level. As an immediate step, the fate of these negative structures on including beyond mean field 
physics needs to be worked out. We also hope that our work motivates a similar study on the LQCD front 
on the sign structures of the off-diagonal susceptibilities which have so far been ignored.

\section{Acknowledgements}
SC acknowledges ``Centre for Nuclear Theory" 
[PIC XII-R\&D-VEC-5.02.0500], Variable Energy Cyclotron Centre for support.
KAM acknowledges the support by the National Science Foundation under Grant No. PHY-0854889. 
The authors would like to thank the organizers of WHEPP-2013 (Workshop on High Energy Particle Physics and Phenomenology) where part of this work was done.
KAM would like to thank K. Sridhar and the Department of Theoretical Physics, TIFR (Mumbai) for hospitality where part of this work was done.

\bibliographystyle{apsrev4-1}
\bibliography{tmusus}

\end{document}